\documentclass[aps,prl,reprint,groupedaddress]{revtex4-2}
\usepackage{dynkin-diagrams}
\usepackage{xfrac}
\usepackage{enumitem}

\DeclareMathSymbol{\shortminus}{\mathbin}{AMSa}{"39}
\newcommand{\dd}{\mathrm{d}}

\newcommand{\Rt}{\widetilde{R}}
\newcommand{\id}{\mathbf{1}}

\newcommand{\ada}{\mathbf{a}}
\newcommand{\adb}{\mathbf{b}}
\newcommand{\adc}{\mathbf{c}}

\newcommand{\Ac}{\mathcal{A}}
\newcommand{\Bc}{\mathcal{B}}

\newcommand{\Eh}{\widehat{E}}
\newcommand{\Et}{\widetilde{E}}
\newcommand{\Mt}{\widetilde{M}}
\newcommand{\Vt}{\widetilde{V}}
\newcommand{\vt}{\widetilde{v}}

\newcommand{\xih}{\widehat{\xi}}
\newcommand{\deltaE}{\delta\mathrm{E}}
\newcommand{\Thetab}{\overline{\Theta}}

\newcommand{\genLie}{\mathbb{L}}

\newcommand{\GL}[1][n]{\mathrm{GL}(#1)}
\newcommand{\GS}{G_{\mathrm{S}}}
\newcommand{\gs}{\mathfrak{g}_{\mathrm{S}}}
\newcommand{\GD}{G_{\mathrm{D}}}
\newcommand{\GM}{G_{\mathrm{M}}}
\newcommand{\Edd}[1][d]{\mathrm{E}_{#1(#1)}}
\newcommand{\Odd}[1][d]{\mathrm{O}(#1,#1)}
\newcommand{\irrep}[1]{\mathbf{#1}}
\newcommand{\irrepb}[1]{\mathbf{\overline{#1}}}

\newcommand{\Calpha}{\boldsymbol{\alpha}}
\newcommand{\Cbeta}{\boldsymbol{\beta}}
\newcommand{\Cgamma}{\boldsymbol{\gamma}}
\newcommand{\Cdelta}{\boldsymbol{\delta}}

\renewcommand{\AA}{\overline{A}}
\newcommand{\BB}{\overline{B}}
\newcommand{\CC}{\overline{C}}
\newcommand{\DD}{\overline{D}}
\newcommand{\EE}{\overline{E}}
\newcommand{\AAA}{\overline{\AA}}
\newcommand{\BBB}{\overline{\BB}}
\newcommand{\CCC}{\overline{\CC}}
\newcommand{\DDD}{\overline{\DD}}

\newcommand{\PS}{Pol\'a\v{c}ek-Siegel }

\begin{document}
\title{The Hierarchy of Curvatures in Exceptional Geometry}

\author{Falk Hassler}
\email[]{falk.hassler@uwr.edu.pl}
\affiliation{University of Wroc\l{}aw, Faculty of Physics and Astronomy, Maksa Borna 9, 50-204 Wroclaw, Poland}
%%%
\author{Yuho Sakatani}
\email[]{yuho@koto.kpu-m.ac.jp}
\affiliation{Department of Physics, Kyoto Prefectural University of Medicine, 1-5 Shimogamohangi-cho, Sakyo-ku, Kyoto 606-0823, Japan}

\date{\today}

\begin{abstract}
  Despite remarkable success in describing supergravity reductions and backgrounds, generalized geometry and the closely related exceptional field theory are still lacking a fundamental object of differential geometry, the Riemann tensor. We explain that to construct such a tensor, an as of yet overlooked hierarchy of connections is required. They complement the spin connection with higher representations known from the tensor hierarchy of gauged supergravities. In addition to solving an important conceptual problem, this idea allows to define and explicitly construct generalized homogeneous spaces. They are the underlying structures of generalized U-duality, admit consistent truncations and provide a huge class of new backgrounds for flux compactifications with non-trivial generalized structure groups.
\end{abstract}


\maketitle

Generalized geometry has evolved to a pivotal tool in dimensional reductions of supergravity with non-vanishing fluxes. Many of its key results have their root in differential geometry adapted to an extended tangent space incorporating forms in addition to a vector. We drop the prefix ``generalized'' in the following and assume it by default. Most mileage is gained when supersymmetry is at least partially preserved. Prominent examples include Scherk-Schwarz reductions to maximal gauged supergravities in two or more dimensions and special structure manifolds that provide the analogs of Calabi-Yau, K\"ahler, and hyper-K\"ahler manifolds in flux compactifications.

Despite all success, a central concept of differential geometry, Riemann curvature, is still lacking a complete understanding in generalized geometry. It is possible to define metric compatible connections, but imposing vanishing torsion does not fix all their components \cite{Siegel:1993th,*Siegel:1993xq,Coimbra:2011nw,*Coimbra:2011ky,*Coimbra:2012af,Hohm:2011si}. Moreover, the familiar expression for the Riemann tensor is not covariant anymore and has to be modified and projected \cite{Siegel:1993th,*Siegel:1993xq,Coimbra:2011nw,*Coimbra:2011ky,*Coimbra:2012af,Hohm:2011si,Park:2013gaj,Cederwall:2013naa,Aldazabal:2013mya}. Although sufficient to capture supergravity at the leading two-derivative level, the current construction is not completely satisfying. Therefore, we present a novel perspective on curvature in generalized geometry based on two key ideas:
\begin{enumerate}[label={\alph*)}]
  \item Torsion and curvature tensors have to transform covariantly under generalized diffeomorphisms \textit{and} a structure group $\GS$.
  \item Reconciling internal and external diffeomorphisms for dimensional reductions is a crucial step in the construction of gauged supergravities, resulting in a hierarchy of gauge fields know as \textit{tensor hierarchy} \cite{deWit:2008ta,*Riccioni:2009xr}.
\end{enumerate}
Both combined suggest to embed the $d$-dimensional space $M_d$, for which curvatures should be obtained, together with its structure group $\GS$ into a larger space, called the mega-space $M_p$ with $p=d+n$ and $n=\dim \GS$. By construction, this new space is parallelizable and thus equipped with a globally defined frame which gives rise to a unique, curvature-free connection. Similar to Cartan geometry, its torsion decomposes into torsion and curvature on $M_d$. This approach can be understood as a direct construction of the tensor hierarchy where a tower of connection and their curvatures arises on $M_d$. At its bottom, we recover all quantities relevant for two-derivative supergravity. Beyond, we find new covariant tensors with more than two derivatives. In addition to conceptual insights, our approach provides an explicit construction for a new, large class of consistent truncations of supergravity and the intriguing web of dualities that relates them. 

\paragraph{Level Decomposition.} Our starting point is a pair of two Lie groups $\GD=\Edd$ with $d\le6$, governing generalized diffeomorphisms on $M_d$, and GL($n$), in which the structure group $\GS \subset \GD$ is embedded. They will not be studied separately but rather are unified in $\GM=\Edd[p]$ which governs diffeomorphisms on $M_p$. There are two different perspectives one can take on this setting: First, start with $\GM$ and remove the $d$\textsuperscript{th} root in its Dynkin diagram
\begin{equation}\label{eqn:leveldecomp}
  \dynkin[edge length=.75cm,labels*={1,$p$,,,$d-1$,$d$,$d+1$,$p-1$}] E{****.*x*.*}
\end{equation}
to obtain the algebras $\GD$ and $\GL\supset\GS$. Taking any irreducible representation (irrep) of $\GM$ and branching it accordingly, results in a sum of irreps of $\GD\times\GL$. They can be organized according to a grading called level. For our purpose, this top-down approach is not ideal because $\GM$ and the mega-space $M_p$ are only book keeping tools. The protagonist is $M_d$ which is controlled by $\GD$ and $\GS$. Therefore, it is better to start with them and only step by step recover $\GM$.

Their generators will be called $K_\ada$ and $K_\alpha^\beta$ for $\GL$ with $\alpha,\beta,\dots\in\{1,\dots,n\}$ and $\ada$ valued in the adjoint representation of $\GD$. They satisfy the non-trivial commutation relations
\begin{equation}
  \begin{aligned}
    [ K_\ada, K_\adb ] &= f_{\ada\adb}{}^{\adc} K_{\adc}\,,\\
    [ K_\alpha^\beta, K_\gamma^\delta ] &= \delta_\alpha^\delta K_\gamma^{\beta} - \delta_\gamma^\beta K_\alpha^\delta\,.
  \end{aligned}
\end{equation}
This completes all level zero contributions in the decomposition of $\GM$'s adjoint representation. To proceed to levels $\mp1$, consider $R^A_\alpha$, with the index $A$ transforming in the $R_1$ representation of $\GD$, and its dual $\Rt_B^\beta$. Their commutation relation with all $K$s are easily fixed by representation theory. More complicated is the commutator
\begin{equation}\label{eqn:commRRt}
  [ \Rt^\alpha_A, R_\beta^B ] = \delta_A^B ( \Cbeta \delta_\alpha^\beta L - K^\alpha_\beta ) + \Calpha (t^{\ada})_A^B K_{\ada}\,,
\end{equation}
where $(t_{\ada})_B^C$ denotes the generators of $\GD$ in the $R_1$ representation. Note that we raise and lower adjoint indices with the Killing metric. Moreover, $L=K_\alpha^\alpha$ is distinguished because its eigenvalues are the levels of the decomposition. Finally, we use the constants $\Calpha$ and $\Cbeta$ from the definition of $\GD$ generalized diffeomorphisms \cite{Berman:2012vc} whose values are listed in the supplementary material. By a rescaling of $\Rt^\alpha_A$ and $R_\beta^B$, it is always possible to fix the coefficient in front of $K_\alpha^\beta$ to minus one. The other two coefficients have to be as given to define the level 2 generators by
\begin{equation}\label{eqn:defR2}
  \begin{aligned}
    [ R^A_\alpha, R^B_\beta ] &= \eta^{AB\CC} R_{\alpha\beta\CC}
      \quad\text{and}\quad \\
    [ \Rt_A^\alpha, \Rt_B^\beta ] &= \eta_{AB\CC} \Rt^{\alpha\beta\CC}
  \end{aligned}
\end{equation}
with the $\eta$-tensors representing the ``square root'' of $\GD$'s $Y$-tensor
\begin{equation}
  \begin{aligned}
    Y^{AB}_{CD} &= \eta^{AB\EE} \eta_{CD\EE} \\
                & = - \Calpha (t^{\ada})_D^A (t_{\ada})_C^B + \Cbeta \delta_D^A \delta_C^B + \delta_C^A\delta_D^B\,.
  \end{aligned}
\end{equation}
This implies that the bared indices label the $R_2$ representation of $\GD$. All other new commutators at this level arise through the Jacobi identity from the lower levels. One can repeat this procedure level by level to obtain generators in higher representations of the tensor hierarchy algebra. However, already at level 2 one sees all relevant features of the construction. Therefore, we stop here and refer to supplementary material for level 3 or the companion article \cite{FULLPAPER} for technical details.

For these generators, we introduce a representation building on the highest weight states $|^\alpha\rangle$. They are annihilated by all $\Rt$ generators and by $K_{\ada}$, while $K_\alpha^\beta$ acts as
\begin{equation}
  K_\alpha^\beta |^\gamma\rangle = \delta_\alpha^\gamma |^\beta\rangle + \Cbeta' \delta_\alpha^\beta |^\gamma\rangle\,,
\end{equation}
where $\Cbeta'$ is now taken with respect to $\GM$ instead of $\GD$. Acting with any $R$ generator(s) produces descendants. For our purpose, only
\begin{equation}
    |^A\rangle = \tfrac1{n} R^A_\alpha |^\alpha\rangle\,,
\end{equation}
and the dual states defined by
\begin{equation}
  \langle_\alpha |^\beta \rangle = \delta_\alpha^\beta\,, \qquad
  \langle_A |^B \rangle = \delta_A^B
\end{equation}
are needed. Using the commutator \eqref{eqn:commRRt}, one can easily show that
\begin{equation}
  \langle_A | = -\tfrac1{n} \langle_\alpha | \Rt^\alpha_A
\end{equation}
has exactly the desired properties.

\paragraph{Exceptional \PS Form.} Next, we fix the form of the frame on $M_p$ to compute the torsion of the flat derivative it induces. Because the lowest levels will be sufficient, it is convenient to suppress $\GM$-indices and instead just use the index-free form
\begin{equation}\label{eqn:decompEh}
  \Eh = \Mt N \Vt
\end{equation}
of the frame. Its splitting is inspired by results for the duality group $\Odd$ \cite{Butter:2022iza} and will be motivated in the following. Finally, we need the generalized Lie derivative \cite{Bossard:2017aae}
\begin{equation}\label{eqn:genLie-mega}
  \genLie_{\langle U|} \langle V| = \langle U|\partial_V\rangle \langle V| + \langle V|\langle U| Z |\partial_U\rangle
\end{equation}
that governs generalized diffeomorphisms on the mega-space $M_p$ with $p\le7$ through the $Z$-tensor
\begin{equation}
  \begin{aligned}
    & Z = - \Calpha K_{\ada} \otimes K^{\ada} + \Cbeta L \otimes L + \Cbeta' \id \otimes \id - K_\alpha^\beta \otimes K_\beta^\alpha \\
    & + R_\alpha^A \otimes \Rt^\alpha_A + \tfrac12 R_{\alpha_1\alpha_2\AA} \otimes \Rt^{\alpha_1\alpha_2\AA} + \dots + (R \leftrightarrow \Rt)\,.
  \end{aligned}
\end{equation}
With $\dots$, we denote suppressed levels higher than two. Our index-free notation in \eqref{eqn:genLie-mega} assumes that $\partial_V$ is the partial derivative acting on $\langle V|$ and the convention $\langle U_1|\langle U_2| A \otimes B |V_2\rangle |V_1\rangle = \langle U_2| B |V_2\rangle\langle U_1| A |V_1\rangle$.

The mega-space's torsion is twisted by $\Mt^{-1}$ to get rid of any dependence on the auxiliary coordinates that $\GS$ introduces, resulting in
\begin{equation}
  X_{\Ac} = \langle_{\Ac} |N|^{\Bc}\rangle\Theta_{\Bc} + \langle_{\Ac}| \Theta_{\Bc} Z N |^{\Bc}\rangle
\end{equation}
with $X_{\Ac} = \begin{pmatrix} X_\alpha & X_A \end{pmatrix}$ and the corresponding Maurer-Cartan form
\begin{equation}\label{eqn:ThetaA}
  \Theta_{\Ac} = \Mt^{-1} D_{\Ac} \Eh \Eh^{-1} \Mt \quad \text{where} \quad \Vt|\partial\rangle = |^{\Ac}\rangle D_{\Ac}\,.
\end{equation}
To recover the tensor hierarchy, we require that $\Eh$ is only generated by generators with zero or negative levels. This renders it an element of a parabolic subgroup of $\GM$. $\Mt$ is not further constrained, while $N$ is nilpotent and $\Vt$ has only level 0 contributions. To see how to choose $\Mt$ and $\Vt$, we first single out the $n$ generators \footnote{It is possible to add negative level generators. Due to the Jacobi identity \eqref{eqn:JacobiLieGS}, their structure constants have to be one-cocycles of $\GS$'s Chevalley–Eilenberg complex $\Omega^\bullet(\gs,V_i)$ in appropriate representations $V_i$. If they are additionally coboundaries, they can be removed by a constant $\GD$-transformation. Therefore only non-trivial elements of the Lie algebra cohomology $H^1(\gs,V_i)$ are relevant. For semi-simple $\GS$, there are no such elements according to Whitehead's lemma. Therefore, we set them here to zero.}
\begin{equation}
  t_\alpha = X_{\alpha\beta}{}^\gamma K_\gamma^\beta + X_\alpha{}^{\adb} K_{\adb}
\end{equation}
(with constant coefficients $X_{\alpha\beta}{}^\gamma$ and $X_{\alpha}{}^{\adb}$) of the parabolic subgroup and impose
\begin{equation}\label{eqn:Xalpha=talpha}
  X_\alpha = t_\alpha\,.
\end{equation}
This requires at least $\langle_\alpha | \Theta_B = 0$. Additionally, we restrict the discussion to the special case $X_{\alpha\beta}{}^{\beta}=0$ for the sake of brevity and find  that \eqref{eqn:Xalpha=talpha} further requires
\begin{equation}
  \Theta_{\alpha} = t_\alpha - \tfrac12 \left( X_{\alpha\beta}{}^\gamma + S_{(\alpha\beta)}{}^\gamma \right) N K_\gamma^\beta N^{-1}\,.
\end{equation}
The symmetric $S_{(\alpha\beta)}{}^\gamma$ will not contribute to $X_A$ and leaves an antisymmetric $X_{\alpha\beta}{}^\gamma$. For $D_\alpha N = 0$ and $D_A \Mt = 0$ it is possible to choose $\Mt$, $N$ and $\Vt$ such that this relation follows from \eqref{eqn:ThetaA}. In this case, we identify
\begin{equation}\label{eqn:MCeqsMtVt}
  \begin{aligned}
    \Mt^{-1} D_\alpha \Mt &= t_\alpha \quad \text{and} \\
    D_\alpha \Vt \Vt^{-1} &= -\tfrac12 \left( X_{\alpha\beta}{}^\gamma + S_{(\alpha\beta)}{}^\gamma \right) K_\gamma^\beta\,.
  \end{aligned}
\end{equation}
In principle one could try to make other identifications. But these two are distinguished because they can be integrated, if (and only if)
\begin{equation}\label{eqn:JacobiLieGS}
  [ t_\alpha, t_\beta ] = X_{\alpha\beta}{}^\gamma t_\gamma
\end{equation}
holds. An immediate consequence is that all $t_\alpha$ generate an $n$-dimensional Lie group, the structure group $\GS$! For consistency, we also verify that $\GS$'s coordinates satisfy the section condition on the mega-space,
\begin{equation}
  Y |^\alpha\rangle |^\beta\rangle D_\alpha \,\cdot\, D_\beta \,\cdot\, = 0\,,
\end{equation}
where $Y = Z + \sigma$ and $\sigma |V_2\rangle |V_1\rangle = |V_1\rangle |V_2\rangle$.

To completely fix $\Eh$, we parameterize the nilpotent part $N$ in the decomposition \eqref{eqn:decompEh} by \footnote{In this form, one can easily compute Maurer-Cartan forms and adjoint actions: Assume first $N = e^{\nu}$. Now one can use the identities
\begin{equation}
  \begin{aligned}
    \dd N N^{-1} &= \sum_{m=0}^\infty \frac{[ \nu, \dd\nu ]_m}{(m+1)!}
      \quad\text{and}\quad \\
    N X N^{-1}   &= \sum_{m=0}^\infty \frac{[ \nu, X ]_m}{m!}\,,
  \end{aligned}
\end{equation}
where the bracket $[X, Y]_m$ is defined for $m\ge 0$ as
\begin{equation}
  [X, Y]_m = [ X, [X, Y]_{m-1} ] \quad\text{with}\quad [X, Y]_0 = Y
\end{equation}}
\begin{equation}
  N = \exp \left( \Omega^\alpha_A R^A_\alpha + \tfrac12 \rho^{\alpha\beta\CC} R_{\alpha\beta\CC} + \dots \right) 
\end{equation}
and impose
\begin{equation}
  D_A \Vt \Vt^{-1} = E_A{}^I \partial_I E E^{-1} = W_A
\end{equation}
which implies with \eqref{eqn:MCeqsMtVt} that $\Vt$ decomposes into $E\in\GD\times\mathbb{R}^+$, the frame on $M_d$, and the left-invariant vectors fields $\vt$ on $\GS$. The last two equations contain only fields restricted to $M_d$ and thereby do not depend on the additional coordinates of $M_p$ introduced by $\GS$. A frame of a similar form was first introduced in \cite{Polacek:2013nla} for $\Odd$ and later further refined to arbitrary structure groups \cite{Butter:2021dtu}. Therefore, we refer to it as the \textit{exceptional \PS form}.

\paragraph{Torsions and Curvatures.} In order to find covariant torsion and curvature tensors, we compute all contributions to $X_A$ at levels less or equal to zero. First, we obtain
\begin{equation}\label{eqn:TABC}
  \begin{aligned}
    T_{AB}{}^C = \langle_B| X_A |^C\rangle = (W+\Omega)_{AB}{}^C + Z^{CE}_{BD} \left( W + \Omega\right)_{EA}{}^D,
  \end{aligned}
\end{equation}
after defining
\begin{equation}
  W_{AB}{}^C = \langle_B|W_A|^C\rangle \quad \text{and} \quad
  \Omega_{AB}{}^C = \Omega_A^\alpha X_{\alpha B}{}^C
\end{equation}
with $X_{\alpha B}{}^C = X_\alpha{}^{\adb} (t_{\adb})_B{}^C$. One can easily check that this is the torsion for the covariant derivative
\begin{equation}\label{eqn:defcovd}
  \nabla_A E_B{}^I = E_A{}^J\partial_J E_B{}^I + \Omega_{AB}{}^C E_C{}^I - E_A{}^J \Gamma_{JK}{}^I E_B{}^K = 0\,.
\end{equation}
Rewritten in curved indices,
\begin{equation}
  T_{IJ}{}^K = \Gamma_{IJ}{}^K + Z^{KM}_{JL} \Gamma_{MI}{}^L\,,
\end{equation}
it reproduces the known expression (3.5) in \cite{Cederwall:2013naa}.

For contributions from negative levels, we schematically write
\begin{equation}\label{eqn:XAneg}
  X_A = \dots + X_{AB}^{\beta} R_\beta^B + \tfrac12 X_A^{\beta_1\beta_2\BB} R_{\beta_1\beta_2\BB} + \dots \,.
\end{equation}
First, consider the $-1$ part, $X_{AB}^\beta$: From the Jacobi identity \eqref{eqn:JacobiLieGS}, it follows that the constants $X_{\alpha B}{}^C$ have to form a representation of the Lie algebra $\gs$ on the generalized tangent space of $M_d$. We impose that this representation is faithful, because otherwise it is not possible to interpret $\GS$ as the structure group on $M_d$. Therefore, the tensor \begin{equation}\label{eqn:R1fromX}
  R_{ABC}{}^D := X^\beta_{AB} X_{\beta C}{}^D
\end{equation}
captures $X^\alpha_{AB}$ completely. Evaluating the latter gives rise to
\begin{equation}\label{eqn:R1}
  \begin{aligned}
    R_{ABC}{}^D &= \rho^{ED}_{AC;BE} + 2 D_{[A} \Omega_{B]C}{}^D  + 2 \Omega_{[A|C}{}^E \Omega_{|B]E}{}^D + \\
    & \left( 2 \Omega_{[AB]}{}^E + \tfrac12 Y^{FE}_{BG} \Omega_{FA}{}^G - T_{AB}{}^E \right) \Omega_{EC}{}^D
  \end{aligned}
\end{equation}
with
\begin{equation}
  \rho^{AB}_{CD;EF} = \rho^{\alpha\beta\CC} X_{\alpha C}{}^A X_{\beta D}{}^B \eta_{EF\CC} \,.
\end{equation}
This is the natural Riemann tensor in generalized geometry. Projecting it, such that $\rho^{ED}_{AC;BE}$ drops out, reproduce the existing results in the literature. Depending on the $R_2$ representation, there are two different common projection: 
\begin{itemize}
  \item $R_2=\mathbf{1}$:  $\GD$ has an invariant metric, $\eta_{AB}$, which is applied to lower the last index of $R_{ABC}{}^D$. After symmetrized with respect to $(AB)\leftrightarrow(CD)$, the generalized Riemann tensor of double field theory \cite{Hohm:2011si} arises.
  \item Otherwise: The indices $B$ and $D$ are contracted and the remaining indices are symmetrizes. Rewriting the result in curved indices with the affine connection recovers equation (5.16) of \cite{Aldazabal:2013mya} or (5.4) of \cite{Cederwall:2013naa}.
\end{itemize}
There is however no need for any projection because $R_{ABC}{}^D$ transforms covariantly under $\GD$ and $\GS$ after taking into account the correct transformation of $\rho$ in \eqref{eqn:trrho}. A similar pattern already holds at the level of the torsion \eqref{eqn:TABC}. With only the frame, it is not covariant unless one introduces the spin-connection $\Omega_{AB}{}^C$ that transforms accordingly. Still, there are projections of $T_{AB}{}^C$ where the latter drops out. They give rise to the intrinsic torsion \cite{Coimbra:2014uxa}, used for example in consistent truncations. Here this pattern repeats after the substitutions $E\rightarrow\Omega\rightarrow\rho$.

Next, we need a field strength for $\rho^{AB}_{CD;EF}$. If follows from \eqref{eqn:XAneg} along the same lines as \eqref{eqn:R1fromX}, namely
\begin{equation}
  R_A{}^{BC}_{DE;FG} = X^{\beta_1\beta_2\BB}_A X_{\beta_1 D}{}^B X_{\beta_2 E}{}^C \eta_{FG\BB}\,,
\end{equation}
and evaluates to\footnote{This antisymmetrization weights each term with 1/2.}
\begin{widetext}
  \begin{equation}\label{eqn:R2}
    \begin{aligned}
      R_A{}^{BC}_{DE;FG} = & \rho^{HBC}_{ADE;FGH} + D_A \rho^{BC}_{DE;FG} - 2 F_{A(F|}{}^H \rho^{BC}_{DE;|G)H} + 2 \rho^{BC}_{HE;FG} \Omega_{AD}{}^H + 2 \rho^{CH}_{ED;FG} \Omega_{AH}{}^B + \rho^{CH}_{EA;FG} \Omega_{HD}{}^B \\
                           & - Y^{HI}_{FG} \left[ R_{AHE}{}^C + D_H \Omega_{AE}{}^{C} + \tfrac13 \Omega_{HA}{}^J \Omega_{JE}{}^C \right] 
                           \Omega_{ID}{}^B - \tfrac16 Y^{HI}_{JK} Y^{JL}_{FG} \Omega_{IA}{}^K \Omega_{LE}{}^C \Omega_{HD}{}^B - \left( {}^A_C \leftrightarrow {}^B_D\right)
    \end{aligned}
  \end{equation}
\end{widetext}
with the fluxes
\begin{equation}
  F_{AB}{}^C = W_{AB}{}^C + Z^{CD}_{EB} W_{DA}{}^E \,.
\end{equation}
Here, we encounter the level 3 connection $\rho^{ABC}_{DEF;GHI}$ which is required to render this new curvature covariant.

\paragraph{Gauge Transformations.} To eventually prove covariance of the derived new torsion/curvatures, we  need the transformation of the various connections under $\GD$-diffeomorphisms and $\GS$. They arise from distinguished $\GM$-diffeomorphisms, mediated by $\genLie_{\langle \xih|} \Eh$, which preserves the \PS form. It is convenient to twist infinitesimal variations of the frame by
\begin{equation}\label{eqn:deltaE}
  \deltaE := N^{-1} \Mt^{-1} (\delta\Eh) \Vt^{-1} = \delta E E^{-1} + N^{-1} \delta N
\end{equation}
to better identify the corresponding shift in its components. We consider mega-space  diffeomorphisms which are parameterized by
\begin{equation}
  \langle \xih | = \xi^\alpha \langle_\alpha| \Vt + \xi^A \langle_A | \Vt\,.
\end{equation}
Note that both, $\xi^\alpha$ and $\xi^A$, are chosen such that they do not depend on the auxiliary coordinates of $M_p$ to find the shift of the frame
\begin{equation}\label{eqn:deltaEfromGS}
  \deltaE = \xi^{\Ac} \Thetab_{\Ac} + \langle \xi | D_{\Ac} \Vt \Vt^{-1} Z |^{\Ac}\rangle + D_{\Ac} \xi^{\Bc} \langle_{\Bc}|Z|^{\Ac}\rangle
\end{equation}
with
\begin{equation}
  \Thetab_{\Ac} = N^{-1} \Mt^{-1} D_{\Ac} \Mt N + N^{-1} D_{\Ac} N + D_{\Ac} \Vt \Vt^{-1}\,.
\end{equation}
By comparing the LHS of \eqref{eqn:deltaEfromGS} with the second line of \eqref{eqn:deltaE}, we read off
\begin{equation}
  \delta E_A{}^I = \genLie_\xi E_A{}^I + \xi_A{}^B E_B{}^I\,.
\end{equation}
Here, the Lie derivative is restricted to $M_D$ and has the parameter $\xi^I = \xi^A E_A{}^I$. For this transformation, we recognize $\xi^A$ as the parameter of $\GD$-diffeomorphisms and $\xi_B{}^C = \xi^\alpha X_{\alpha B}{}^C$ as generator of $\GS$-transformations. Any covariant tensor should transform in the same way. We therefore define the anomalous part of the transformation as
\begin{equation}
  \Delta_\xi = \delta - \genLie_\xi - \xi \cdot\,,
\end{equation}
where we understand the last contribution as the action of $\xi_A{}^B$ on each index of the tensor under consideration.

In contrast to the frame, all other connections do not transform covariantly. Instead, we find
\begin{equation}
  \Delta_\xi \Omega_{AB}{}^C = - D_A \xi_B{}^C
\end{equation}
after taking into account $\delta X_{\alpha B}{}^C = 0$. Again this result is in perfect agreement with the literature, because it implies (remember $[\Delta_\xi, \nabla_A] = 0$) 
\begin{equation}
  \Delta_\xi \Gamma_{IJ}{}^K = Z^{KL}_{MJ} \partial_I\partial_L \xi^M 
\end{equation}
and thus matches (3.4) of \cite{Cederwall:2013naa}. On the next level, we obtain
\begin{equation}\label{eqn:trrho}
  \Delta_\xi \rho^{AB}_{CD;EF} = Y^{GH}_{EF} D_G \xi_C{}^A \Omega_{HD}{}^B - \left({}^A_C \leftrightarrow {}^B_D\right)\,.
\end{equation}
One can now check that the expression for the Riemann tensor in \eqref{eqn:R1} satisfies $\Delta_\xi R_{ABCD}=0$ and therewith is covariantly. In the same vein, one finds that the curvature of $\rho^{AB}_{CD;EF}$ in general is only covariant after introducing the level 3 curvature in \eqref{eqn:R2}. This pattern continues until the level decomposition of $\GM$'s adjoint has reached the ``top'' curvature that is covariant on its own.

A central element in the construction is the mega-space Lie derivative \eqref{eqn:genLie-mega} that requires $p\le 7$. Remarkably, one still finds that with the transformations given here the derived expression from above transform covariantly without any restrictions on $p$.

\paragraph{Equivaraint Frames.} It is instructive to flip the perspective and ask: What are the constraints on an arbitrary mega-space frame to admit a \PS form after an appropriate transformation? Clearly, it has to contain $n$ vectors $\langle k_\alpha|$, that generate the Lie algebra $\gs$ through
\begin{equation}\label{eqn:genLie->LieGs}
  \genLie_{\langle k_\alpha|} \langle k_\beta| = f_{\alpha\beta}{}^\gamma \langle k_\gamma|\,.
\end{equation}
Furthermore, the section condition has to be satisfied on the mega-space. It introduces a second level decomposition with respect to the $p$\textsuperscript{th} root in \eqref{eqn:leveldecomp} and break $\Edd[p]$ further down to $\GL[d(-1)]\times\GL$ for M-theory(type IIB). We denote indices enumerating the fundamental of the first group in this product by lowercase Latin letters and the grading follows directly from their position: Each up/down index contributes with $+1$/$-1$. Considering this grading, a natural parameterization of the mega-space frame is
\begin{equation}
  \Eh = L \Et U\,.
\end{equation}
All negative elements produce the matrix $L$, followed by a diagonal $\Et$ that originates from all level 0 generators and finally an upper-triangular matrix $U$ from the rest. Either $L$ or $U$ can be removed by acting with the maximal compact subgroup of $\GM$ from the left. In the context of supergravity, usually $L$ is eliminated while $U$ captures all form-fields and $\Et$ the frame on $M_p$.

For frames in \PS form any contributions from positive level generators (with respect to the decomposition \eqref{eqn:leveldecomp}) have to vanish. $L$ is not affected by them and stays unconstrained. The frame $e$ that governs $\Et$ has the components $e_\alpha{}^\mu$, $e_\alpha{}^i$, $e_a{}^\mu$ and $e_a{}^i$, but to avoid $\Rt^\alpha_a$-contributions, $e_\alpha{}^i = 0$ has to hold. A consequence is that $e_\alpha{}^\mu$ is invertible. Finally, $U$ has to satisfy
\begin{equation}\label{eqn:fluxConditionPS}
  \langle_\alpha| U = 0\,.
\end{equation}
In the M-theory section, $U$ is parameterized in terms of a three-form $C^{(3)}$ and a six-form $C^{(6)}$ on $M_p$. Here \eqref{eqn:fluxConditionPS} for example implies
\begin{equation}
  \iota_\alpha C^{(3)} = 0 \quad\text{and}\quad \iota_\alpha C^{(6)}  = 0\,,
\end{equation}
where $\iota_\alpha$ denotes the interior product with respect to the vector field $e_\alpha{}^\mu \partial_\mu$. This situation can be always achieved by an appropriate coordinate change and a form-field gauge transformations if \eqref{eqn:genLie->LieGs} holds, resulting in $\langle k_\alpha | = \langle_\alpha | \Eh = \langle_\alpha| \Et$\,. Last but not least, one has to verify that the action of $\langle k_\alpha|$ extends to the full frame by
\begin{equation}
  \genLie_{\langle k_\alpha|} \Eh \Eh^{-1} = t_\alpha
\end{equation}
with a constant right-hand side. We call frames with this property \textit{equivariant frames}. 

\paragraph{Exceptional Generalized Cosets.} An important class of these frames arise from generalized parallelizations \cite{Grana:2008yw,*Lee:2014mla}. The latter are constructed on a coset $M_p=H \backslash G$ \cite{Hassler:2022egz} and play a crucial role in the construction of maximally gauged supergravities by consistent truncations. In order to solve the section condition, $H$ has to be a co-isotropic subgroup of $G$. By choosing a second isotropic subgroup $\GS \subset G$, one obtains the scenario discussed above with $M_d=H\backslash G / \GS$ being a double coset. Furthermore, the embedding tensor on the mega-space $X_{\Ac}$ is constant and invariant under the action of $\GS$. Hence, all torsions and curvatures, we computed are covariantly constant with respect to $\nabla_A$.

According to Ambrose and Singer, structure compatible connections with covariantly constant torsion and curvature are in one-to-one correspondence with homogenous spaces \cite{Ambrose1958}. Here, we encounter the lift of this idea to generalized geometry and thus call the cosets $M_d=H\backslash G / \GS$ \textit{exceptional generalized homogenous spaces}. They come with some remarkable properties:
\begin{enumerate}[label={\alph*)}]
  \item\label{item:singular} The $\GS$-action can have fixed points which result in singularities on $M_d$. They might hint towards additional localized objects which are well known in the context of supergravity as branes and monopoles.
  \item In general, there are different admissible co-isotropic subgroups $H$. While each of them leads to a different space and frame, they share the same torsion and curvatures. For the duality group $\Odd$ this phenomena is known as generalized T-duality \cite{Klimcik:1996np,*Demulder:2019vvh,*Sakatani:2021skx}. Here it becomes \textit{generalized U-duality}.
  \item\label{item:consistent} Their intrinsic torsion is constant and a singlet under the action of $\GS$. Therefore, they admit consistent truncations according to theorem 2 of \cite{Cassani:2019vcl}.
  \item\label{item:higherderiv} Because the frame on the mega-space is completely fixed by its embedding tensor, all components of the connections are determined. At the same time, all higher-level curvatures are in general non-trivial.
\end{enumerate}
Items \ref{item:singular}-\ref{item:consistent} makes them perfectly suited as backgrounds for dimensional reductions with various applications in flux compactifications and gauged supergravities. While \ref{item:higherderiv} provides an ideal testing ground to address still open challenges in generalized geometry, like undetermined connection/curvature components and higher-derivative corrections.
 
\begin{acknowledgments}
\paragraph{Acknowledgments.} We would like to thank Martin Cederwall, Achilles Gitsis, Ond\v{r}ej Hul\'ik, Eric Lescano and Jakob Palmkvist for discussions and David Osten and Alex Swash for comments on the draft. FH is supported by the SONATA BIS grant 2021/42/E/ST2/00304 from the National Science Centre (NCN),
Poland and is grateful to the Mainz Institute for Theoretical Physics (MITP) of the DFG Cluster of Excellence PRISMA+ (Project ID 39083149), for its hospitality and its partial support during the completion of parts of this work. YS is supported by JSPS KAKENHI Grant Number JP23K03391.
\end{acknowledgments}

\appendix
\section{Supplementary material}\label{app:algebra}
For the reader's convenience table~\ref{tab:constants} summarizes important constants and representation for duality groups relevant to this letter. Moreover, we provide for completeness all commutators that are required to compute the torsion and curvatures in the main text explicitly.

Lets start with the level $\pm3$ generators
\begin{equation}
  \begin{aligned}
    \left[ [ R^A_{[\alpha}, R^B_\beta ], R^C_{\gamma]} \right] &= \eta^{ABC\DDD} R_{\alpha\beta\gamma\DDD} \quad\text{and} \\
    \left[ [ \Rt_A^{[\alpha}, \Rt_B^\beta ], \Rt_C^{\gamma]} \right] &= \eta_{ABC\DDD} \Rt^{\alpha\beta\gamma\DDD}
  \end{aligned}
\end{equation}
with double bared indices in the $R_3$ representation. Taking into account the antisymmetry with respect to the indices $\alpha$, $\beta$ and $\gamma$, the Jacobi identity in combination with \eqref{eqn:defR2} implies $\eta^{(AB)C\DDD} = \eta^{ABC\DDD}$ and $\eta^{(ABC)\DDD} = 0$. Of course the same holds for the version with lowered indices. For $d\le5$ this observation is sufficient to completely fix $\eta^{ABC\DDD}$ up to a factor because $(R_1\otimes R_2) \cap \left( R_1^{\otimes 3} \right)_{\mathrm{sym}} = R_3$ \cite{Berman:2012vc}. Finally the normalization is fixed by defining
\begin{equation}
  \begin{aligned}
    Y^{ABC}_{DEF} &= \eta^{ABC\DDD} \eta_{DEF\DDD} \\
                  &= ( Y^{AB}_{DE}\delta^C_F - Y^{AB}_{GF} Y^{GC}_{DE} ) \,.
  \end{aligned}
\end{equation}
The action of all level-zero generators follows directly from the representation theory of $\GD$ and GL($n$). It reads for the levels $\pm1$
\begin{equation}
  \begin{aligned}
    [ K_{\ada}, R_\beta^B ] &= \phantom{-} (t_{\ada})_C^B R_\beta^C\,, & 
    [ K_\alpha^\beta, R_\gamma^C ] &= - \delta_\gamma^\beta R_\alpha^C\,,  \\
    [ K_{\ada}, \Rt^\beta_B ] &= - (t_{\ada})_B^C \Rt^\beta_C\,, &
    [ K_\alpha^\beta, \Rt^\gamma_C ] &= \phantom{-} \delta_\alpha^\gamma \Rt^\beta_C \,,
  \end{aligned}
\end{equation}
for the levels $\pm2$,
\begin{equation}
  \begin{aligned}
    [ K_{\ada}, R_{\beta_1\beta_2\BB}] &= \phantom{-} (t_{\ada})_{\BB}^{\CC} R_{\beta_1\beta_2\CC}\,, \\ 
    [ K_{\ada}, \Rt^{\beta_1\beta_2\BB}] &= - (t_{\ada})_{\CC}^{\BB} \Rt^{\beta_1\beta_2\CC} \,, \\
    [ K_\alpha^\beta, R_{\gamma_1\gamma_2\CC}] &= \phantom{-} 2 \delta_{[\gamma_1}^\beta R_{\gamma_2]\alpha\CC} \,, \\
    [ K_\alpha^\beta,\Rt^{\gamma_1\gamma_2\CC}] &= -2 \delta_\alpha^{[\gamma_1} \Rt^{\gamma_2]\beta\CC}\,,
  \end{aligned}
\end{equation}
and for the levels $\pm3$,
\begin{equation}
  \begin{aligned}
    [ K_{\ada}, R_{\beta_1\beta_2\beta_3\BBB}] &= \phantom{-} (t_{\ada})_{\BBB}^{\CCC} R_{\beta_1\beta_2\beta_3\CCC}\,\\
    [ K_{\ada}, \Rt^{\beta_1\beta_2\beta_3\BBB}] &= - (t_{\ada})_{\CCC}^{\BBB} \Rt^{\beta_1\beta_2\beta_3\CCC}\,, \\
    [ K_\alpha^\beta, R_{\gamma_1\gamma_2\gamma_3\CCC}] &= - 3 \delta_{[\gamma_1}^\beta R_{\gamma_2\gamma_3]\alpha\CCC}\,, \\
    [ K_\alpha^\beta, \Rt^{\gamma_1\gamma_2\gamma_3\CCC}] &= \phantom{-} 3 \delta_\alpha^{[\gamma_1} \Rt^{\gamma_2\gamma_3]\beta\CCC}\,.
  \end{aligned}
\end{equation}
In particular, $(t_{\ada})_C^B$ denotes the generators of the former in the $R_1$ representation, whereas 
\begin{equation}
  \begin{aligned}
    (t_{\ada})_{\BB}^{\CC} &= \tfrac2\Cgamma (t_{\ada})_D^E \eta_{EF\BB} \eta^{DF\CC} \quad \text{and} \\
    (t_{\ada})_{\BBB}^{\CCC} &= \tfrac1\Cdelta (t_{\ada})_D^E \left( 2 \eta_{EFG\BBB}\eta^{DFG\CCC} +\eta_{FGE\BBB}\eta^{FGD\CCC}\right)
  \end{aligned}
\end{equation}
acts on the $R_2$ and $R_3$ representation respectively. Here a new normalization factor, $\Cdelta=2(r-1)r$, appears which can be expressed in terms of the rank $r$ of $\GD$. One also needs the commutators between positive and negative level generators resulting in level $\pm1$ and $\pm2$ contributions. First, the levels $\pm2$ give rise to
\begin{equation}
  \begin{aligned}
    [ \Rt^\alpha_A, R_{\beta_1\beta_2\beta_3\BBB} ] &= - 3 \delta_{[\beta_1} R_{\beta_2\beta_3]\CC} Z_A{}^{\CC}{}_{\BBB}\,, \\
    [ \Rt^{\alpha_1\alpha_2\alpha_3\AAA}, R_\beta^B ] &= \phantom{-} 3 \delta_\beta^{[\alpha_1} \Rt^{\alpha_2\alpha_3]\CC} Z^B{}_{\CC}{}^{\AAA}\,.
  \end{aligned}
\end{equation}
where we defined the intertwiners
\begin{equation}
  \begin{aligned}
    Z_A{}^{\BB}{}_{\CCC} &= \tfrac1{\Cgamma} \eta_{DEA\CCC}\eta^{DE\BB} \quad \text{and} \\
    Z^A{}_{\BB}{}^{\CCC} &= \tfrac1{\Cgamma} \eta^{DEA\CCC}\eta_{DE\BB}\,.
  \end{aligned}
\end{equation}
For the levels $\pm1$, we obtain
\begin{equation}
  \begin{aligned}
    [ \Rt^{\alpha_1\alpha_2\AA}, R_{\beta_1\beta_2\beta_3\BBB}] &= \phantom{-} \delta_{[\beta_1}^{\alpha_1} \delta_{\beta_2}^{\alpha_2} R_{\beta_3]}^C Z_C{}^{\AA}{}_{\BBB}\,, \\
    [ \Rt^{\alpha_1\alpha_2\alpha_3\AAA}, R_{\beta_1\beta_2\BB} ] &= -\delta_{\beta_1}^{[\alpha_1} \delta_{\beta_2}^{\alpha_2} \Rt^{\alpha_3]}_C Z^C{}_{\BB}{}^{\AAA} \,, \\
    [ \Rt^\alpha_A, R_{\beta_1\beta_2\BB} ] &= \phantom{-} \delta_{[\beta_1}^\alpha R_{\beta_2]}^C \eta_{CA\BB} \,, \\
    [ \Rt^{\alpha_1\alpha_2\AA}, R_\beta^B ] &= \phantom{-} \delta_\beta^{[\alpha_1} \Rt^{\alpha_2]}_C \eta^{CB\AA} \,,
  \end{aligned}
\end{equation}
and finally at level 0,
\begin{equation}
  \begin{gathered}
    [ \Rt^{\alpha_1\alpha_2\alpha_3\AAA}, R_{\beta_1\beta_2\beta_3\BBB} ] = 6 \Calpha \delta^{\alpha_1\alpha_2\alpha_3}_{\beta_1\beta_2\beta_3} (t^{\ada})_{\BBB}^{\AAA} K_{\ada}
    \\ + 18 \left( \Cbeta \delta^{\alpha_1\alpha_2\alpha_3}_{\beta_1\beta_2\beta_3} L - \delta_{[\beta_1}^{[\alpha_1} \delta_{\beta_2}^{\alpha_2} K_{\beta_3]}^{\alpha_3]} \right)\delta_{\BBB}^{\AAA} 
  \end{gathered}
\end{equation}
and
\begin{equation}
  \begin{gathered}
    [ \Rt^{\alpha_1\alpha_2\AA}, R_{\beta_1\beta_2\BB} ] = - 2 \Calpha \delta_{\beta_1\beta_2}^{\alpha_1\alpha_2} (t^{\ada})_{\BB}^{\AA} K_{\ada} \\
    + 4 \left( \Cbeta \delta_{\beta_1\beta_2}^{\alpha_1\alpha_2} L - \delta_{[\beta_1}^{[\alpha_1} K_{\beta_2]}^{\alpha_2]} \right) \delta_{\BB}^{\AA}\,.
  \end{gathered}
\end{equation}
All of them arise from already known commutators by the Jacobi identity.
\begin{table}
  \caption{Relevant representations and constants}\label{tab:constants}
  \begin{ruledtabular}
    \begin{tabular}{lcccccc}
      & $\Odd$ & SL(5) & Spin(5,5) & $\Edd[6]$ & $\Edd[7]$\\\hline
      $\Calpha$ & 2 & 3 & 4 & 6 & 12\\
      $\Cbeta$  & 0 & 1/5 & 1/4 & 1/3 & 1/2 \\
      $\Cgamma$ & $2 d$ & 6 & 8 & 10 & --\\
      $r$ (rank) & d & 4 & 5 & 6 & 7 \\
      adj & $\irrep{d(2d-1)}$ & $\irrep{24}$ & $\irrep{45}$ & $\irrep{78}$ & $\irrep{133}$ \\
      $R_1$ & $\irrep{2d}$ & $\irrepb{10}$ & $\irrep{16_{\mathrm{c}}}$ & $\irrep{27}$  & $\irrep{56}$ \\
      $R_2$ & $\irrep{1}$  & $\irrep{5}$   & $\irrep{10}$ & $\irrepb{27}$ & $\irrep{113}$ \\
      $R_3$ & -- & $\irrepb{5}$ & $\irrep{16_{\mathrm{s}}}$ & $\irrep{78}$ & $\irrep{912}$
    % Lines of table here ending with \\
    \end{tabular}
  \end{ruledtabular}
\end{table}

\bibliography{literature.bib}

\begin{thebibliography}{29}%
\makeatletter
\providecommand \@ifxundefined [1]{%
 \@ifx{#1\undefined}
}%
\providecommand \@ifnum [1]{%
 \ifnum #1\expandafter \@firstoftwo
 \else \expandafter \@secondoftwo
 \fi
}%
\providecommand \@ifx [1]{%
 \ifx #1\expandafter \@firstoftwo
 \else \expandafter \@secondoftwo
 \fi
}%
\providecommand \natexlab [1]{#1}%
\providecommand \enquote  [1]{``#1''}%
\providecommand \bibnamefont  [1]{#1}%
\providecommand \bibfnamefont [1]{#1}%
\providecommand \citenamefont [1]{#1}%
\providecommand \href@noop [0]{\@secondoftwo}%
\providecommand \href [0]{\begingroup \@sanitize@url \@href}%
\providecommand \@href[1]{\@@startlink{#1}\@@href}%
\providecommand \@@href[1]{\endgroup#1\@@endlink}%
\providecommand \@sanitize@url [0]{\catcode `\\12\catcode `\$12\catcode `\&12\catcode `\#12\catcode `\^12\catcode `\_12\catcode `\%12\relax}%
\providecommand \@@startlink[1]{}%
\providecommand \@@endlink[0]{}%
\providecommand \url  [0]{\begingroup\@sanitize@url \@url }%
\providecommand \@url [1]{\endgroup\@href {#1}{\urlprefix }}%
\providecommand \urlprefix  [0]{URL }%
\providecommand \Eprint [0]{\href }%
\providecommand \doibase [0]{https://doi.org/}%
\providecommand \selectlanguage [0]{\@gobble}%
\providecommand \bibinfo  [0]{\@secondoftwo}%
\providecommand \bibfield  [0]{\@secondoftwo}%
\providecommand \translation [1]{[#1]}%
\providecommand \BibitemOpen [0]{}%
\providecommand \bibitemStop [0]{}%
\providecommand \bibitemNoStop [0]{.\EOS\space}%
\providecommand \EOS [0]{\spacefactor3000\relax}%
\providecommand \BibitemShut  [1]{\csname bibitem#1\endcsname}%
\let\auto@bib@innerbib\@empty
%</preamble>
\bibitem [{\citenamefont {Siegel}(1993{\natexlab{a}})}]{Siegel:1993th}%
  \BibitemOpen
  \bibfield  {author} {\bibinfo {author} {\bibfnamefont {W.}~\bibnamefont {Siegel}},\ }\bibfield  {title} {\bibinfo {title} {{Superspace duality in low-energy superstrings}},\ }\href {https://doi.org/10.1103/PhysRevD.48.2826} {\bibfield  {journal} {\bibinfo  {journal} {Phys. Rev. D}\ }\textbf {\bibinfo {volume} {48}},\ \bibinfo {pages} {2826} (\bibinfo {year} {1993}{\natexlab{a}})},\ \Eprint {https://arxiv.org/abs/hep-th/9305073} {arXiv:hep-th/9305073} \BibitemShut {NoStop}%
\bibitem [{\citenamefont {Siegel}(1993{\natexlab{b}})}]{Siegel:1993xq}%
  \BibitemOpen
  \bibfield  {author} {\bibinfo {author} {\bibfnamefont {W.}~\bibnamefont {Siegel}},\ }\bibfield  {title} {\bibinfo {title} {{Two vierbein formalism for string inspired axionic gravity}},\ }\href {https://doi.org/10.1103/PhysRevD.47.5453} {\bibfield  {journal} {\bibinfo  {journal} {Phys. Rev. D}\ }\textbf {\bibinfo {volume} {47}},\ \bibinfo {pages} {5453} (\bibinfo {year} {1993}{\natexlab{b}})},\ \Eprint {https://arxiv.org/abs/hep-th/9302036} {arXiv:hep-th/9302036} \BibitemShut {NoStop}%
\bibitem [{\citenamefont {Coimbra}\ \emph {et~al.}(2011)\citenamefont {Coimbra}, \citenamefont {Strickland-Constable},\ and\ \citenamefont {Waldram}}]{Coimbra:2011nw}%
  \BibitemOpen
  \bibfield  {author} {\bibinfo {author} {\bibfnamefont {A.}~\bibnamefont {Coimbra}}, \bibinfo {author} {\bibfnamefont {C.}~\bibnamefont {Strickland-Constable}},\ and\ \bibinfo {author} {\bibfnamefont {D.}~\bibnamefont {Waldram}},\ }\bibfield  {title} {\bibinfo {title} {{Supergravity as Generalised Geometry I: Type II Theories}},\ }\href {https://doi.org/10.1007/JHEP11(2011)091} {\bibfield  {journal} {\bibinfo  {journal} {JHEP}\ }\textbf {\bibinfo {volume} {11}},\ \bibinfo {pages} {091}},\ \Eprint {https://arxiv.org/abs/1107.1733} {arXiv:1107.1733 [hep-th]} \BibitemShut {NoStop}%
\bibitem [{\citenamefont {Coimbra}\ \emph {et~al.}(2014{\natexlab{a}})\citenamefont {Coimbra}, \citenamefont {Strickland-Constable},\ and\ \citenamefont {Waldram}}]{Coimbra:2011ky}%
  \BibitemOpen
  \bibfield  {author} {\bibinfo {author} {\bibfnamefont {A.}~\bibnamefont {Coimbra}}, \bibinfo {author} {\bibfnamefont {C.}~\bibnamefont {Strickland-Constable}},\ and\ \bibinfo {author} {\bibfnamefont {D.}~\bibnamefont {Waldram}},\ }\bibfield  {title} {\bibinfo {title} {{$E_{d(d)} \times \mathbb{R}^+$ generalised geometry, connections and M theory}},\ }\href {https://doi.org/10.1007/JHEP02(2014)054} {\bibfield  {journal} {\bibinfo  {journal} {JHEP}\ }\textbf {\bibinfo {volume} {02}},\ \bibinfo {pages} {054}},\ \Eprint {https://arxiv.org/abs/1112.3989} {arXiv:1112.3989 [hep-th]} \BibitemShut {NoStop}%
\bibitem [{\citenamefont {Coimbra}\ \emph {et~al.}(2014{\natexlab{b}})\citenamefont {Coimbra}, \citenamefont {Strickland-Constable},\ and\ \citenamefont {Waldram}}]{Coimbra:2012af}%
  \BibitemOpen
  \bibfield  {author} {\bibinfo {author} {\bibfnamefont {A.}~\bibnamefont {Coimbra}}, \bibinfo {author} {\bibfnamefont {C.}~\bibnamefont {Strickland-Constable}},\ and\ \bibinfo {author} {\bibfnamefont {D.}~\bibnamefont {Waldram}},\ }\bibfield  {title} {\bibinfo {title} {{Supergravity as Generalised Geometry II: $E_{d(d)} \times \mathbb{R}^+$ and M theory}},\ }\href {https://doi.org/10.1007/JHEP03(2014)019} {\bibfield  {journal} {\bibinfo  {journal} {JHEP}\ }\textbf {\bibinfo {volume} {03}},\ \bibinfo {pages} {019}},\ \Eprint {https://arxiv.org/abs/1212.1586} {arXiv:1212.1586 [hep-th]} \BibitemShut {NoStop}%
\bibitem [{\citenamefont {Hohm}\ and\ \citenamefont {Zwiebach}(2012)}]{Hohm:2011si}%
  \BibitemOpen
  \bibfield  {author} {\bibinfo {author} {\bibfnamefont {O.}~\bibnamefont {Hohm}}\ and\ \bibinfo {author} {\bibfnamefont {B.}~\bibnamefont {Zwiebach}},\ }\bibfield  {title} {\bibinfo {title} {{On the Riemann Tensor in Double Field Theory}},\ }\href {https://doi.org/10.1007/JHEP05(2012)126} {\bibfield  {journal} {\bibinfo  {journal} {JHEP}\ }\textbf {\bibinfo {volume} {05}},\ \bibinfo {pages} {126}},\ \Eprint {https://arxiv.org/abs/1112.5296} {arXiv:1112.5296 [hep-th]} \BibitemShut {NoStop}%
\bibitem [{\citenamefont {Park}\ and\ \citenamefont {Suh}(2013)}]{Park:2013gaj}%
  \BibitemOpen
  \bibfield  {author} {\bibinfo {author} {\bibfnamefont {J.-H.}\ \bibnamefont {Park}}\ and\ \bibinfo {author} {\bibfnamefont {Y.}~\bibnamefont {Suh}},\ }\bibfield  {title} {\bibinfo {title} {{U-geometry: SL(5)}},\ }\href {https://doi.org/10.1007/JHEP04(2013)147} {\bibfield  {journal} {\bibinfo  {journal} {JHEP}\ }\textbf {\bibinfo {volume} {04}},\ \bibinfo {pages} {147}},\ \bibinfo {note} {[Erratum: JHEP 11, 210 (2013)]},\ \Eprint {https://arxiv.org/abs/1302.1652} {arXiv:1302.1652 [hep-th]} \BibitemShut {NoStop}%
\bibitem [{\citenamefont {Cederwall}\ \emph {et~al.}(2013)\citenamefont {Cederwall}, \citenamefont {Edlund},\ and\ \citenamefont {Karlsson}}]{Cederwall:2013naa}%
  \BibitemOpen
  \bibfield  {author} {\bibinfo {author} {\bibfnamefont {M.}~\bibnamefont {Cederwall}}, \bibinfo {author} {\bibfnamefont {J.}~\bibnamefont {Edlund}},\ and\ \bibinfo {author} {\bibfnamefont {A.}~\bibnamefont {Karlsson}},\ }\bibfield  {title} {\bibinfo {title} {{Exceptional geometry and tensor fields}},\ }\href {https://doi.org/10.1007/JHEP07(2013)028} {\bibfield  {journal} {\bibinfo  {journal} {JHEP}\ }\textbf {\bibinfo {volume} {07}},\ \bibinfo {pages} {028}},\ \Eprint {https://arxiv.org/abs/1302.6736} {arXiv:1302.6736 [hep-th]} \BibitemShut {NoStop}%
\bibitem [{\citenamefont {Aldazabal}\ \emph {et~al.}(2013)\citenamefont {Aldazabal}, \citenamefont {Gra\~na}, \citenamefont {Marqu\'es},\ and\ \citenamefont {Rosabal}}]{Aldazabal:2013mya}%
  \BibitemOpen
  \bibfield  {author} {\bibinfo {author} {\bibfnamefont {G.}~\bibnamefont {Aldazabal}}, \bibinfo {author} {\bibfnamefont {M.}~\bibnamefont {Gra\~na}}, \bibinfo {author} {\bibfnamefont {D.}~\bibnamefont {Marqu\'es}},\ and\ \bibinfo {author} {\bibfnamefont {J.~A.}\ \bibnamefont {Rosabal}},\ }\bibfield  {title} {\bibinfo {title} {{Extended geometry and gauged maximal supergravity}},\ }\href {https://doi.org/10.1007/JHEP06(2013)046} {\bibfield  {journal} {\bibinfo  {journal} {JHEP}\ }\textbf {\bibinfo {volume} {06}},\ \bibinfo {pages} {046}},\ \Eprint {https://arxiv.org/abs/1302.5419} {arXiv:1302.5419 [hep-th]} \BibitemShut {NoStop}%
\bibitem [{\citenamefont {de~Wit}\ \emph {et~al.}(2008)\citenamefont {de~Wit}, \citenamefont {Nicolai},\ and\ \citenamefont {Samtleben}}]{deWit:2008ta}%
  \BibitemOpen
  \bibfield  {author} {\bibinfo {author} {\bibfnamefont {B.}~\bibnamefont {de~Wit}}, \bibinfo {author} {\bibfnamefont {H.}~\bibnamefont {Nicolai}},\ and\ \bibinfo {author} {\bibfnamefont {H.}~\bibnamefont {Samtleben}},\ }\bibfield  {title} {\bibinfo {title} {{Gauged Supergravities, Tensor Hierarchies, and M-Theory}},\ }\href {https://doi.org/10.1088/1126-6708/2008/02/044} {\bibfield  {journal} {\bibinfo  {journal} {JHEP}\ }\textbf {\bibinfo {volume} {02}},\ \bibinfo {pages} {044}},\ \Eprint {https://arxiv.org/abs/0801.1294} {arXiv:0801.1294 [hep-th]} \BibitemShut {NoStop}%
\bibitem [{\citenamefont {Riccioni}\ \emph {et~al.}(2009)\citenamefont {Riccioni}, \citenamefont {Steele},\ and\ \citenamefont {West}}]{Riccioni:2009xr}%
  \BibitemOpen
  \bibfield  {author} {\bibinfo {author} {\bibfnamefont {F.}~\bibnamefont {Riccioni}}, \bibinfo {author} {\bibfnamefont {D.}~\bibnamefont {Steele}},\ and\ \bibinfo {author} {\bibfnamefont {P.}~\bibnamefont {West}},\ }\bibfield  {title} {\bibinfo {title} {{The E(11) origin of all maximal supergravities: The Hierarchy of field-strengths}},\ }\href {https://doi.org/10.1088/1126-6708/2009/09/095} {\bibfield  {journal} {\bibinfo  {journal} {JHEP}\ }\textbf {\bibinfo {volume} {09}},\ \bibinfo {pages} {095}},\ \Eprint {https://arxiv.org/abs/0906.1177} {arXiv:0906.1177 [hep-th]} \BibitemShut {NoStop}%
\bibitem [{\citenamefont {Berman}\ \emph {et~al.}(2013)\citenamefont {Berman}, \citenamefont {Cederwall}, \citenamefont {Kleinschmidt},\ and\ \citenamefont {Thompson}}]{Berman:2012vc}%
  \BibitemOpen
  \bibfield  {author} {\bibinfo {author} {\bibfnamefont {D.~S.}\ \bibnamefont {Berman}}, \bibinfo {author} {\bibfnamefont {M.}~\bibnamefont {Cederwall}}, \bibinfo {author} {\bibfnamefont {A.}~\bibnamefont {Kleinschmidt}},\ and\ \bibinfo {author} {\bibfnamefont {D.~C.}\ \bibnamefont {Thompson}},\ }\bibfield  {title} {\bibinfo {title} {{The gauge structure of generalised diffeomorphisms}},\ }\href {https://doi.org/10.1007/JHEP01(2013)064} {\bibfield  {journal} {\bibinfo  {journal} {JHEP}\ }\textbf {\bibinfo {volume} {01}},\ \bibinfo {pages} {064}},\ \Eprint {https://arxiv.org/abs/1208.5884} {arXiv:1208.5884 [hep-th]} \BibitemShut {NoStop}%
\bibitem [{\citenamefont {Hassler}\ and\ \citenamefont {Sakatani}()}]{FULLPAPER}%
  \BibitemOpen
  \bibfield  {author} {\bibinfo {author} {\bibfnamefont {F.}~\bibnamefont {Hassler}}\ and\ \bibinfo {author} {\bibfnamefont {Y.}~\bibnamefont {Sakatani}},\ }\href@noop {} {\bibinfo {title} {{\textit{to appear}}}}\BibitemShut {NoStop}%
\bibitem [{\citenamefont {Butter}\ \emph {et~al.}(2023)\citenamefont {Butter}, \citenamefont {Hassler}, \citenamefont {Pope},\ and\ \citenamefont {Zhang}}]{Butter:2022iza}%
  \BibitemOpen
  \bibfield  {author} {\bibinfo {author} {\bibfnamefont {D.}~\bibnamefont {Butter}}, \bibinfo {author} {\bibfnamefont {F.}~\bibnamefont {Hassler}}, \bibinfo {author} {\bibfnamefont {C.~N.}\ \bibnamefont {Pope}},\ and\ \bibinfo {author} {\bibfnamefont {H.}~\bibnamefont {Zhang}},\ }\bibfield  {title} {\bibinfo {title} {{Consistent truncations and dualities}},\ }\href {https://doi.org/10.1007/JHEP04(2023)007} {\bibfield  {journal} {\bibinfo  {journal} {JHEP}\ }\textbf {\bibinfo {volume} {04}},\ \bibinfo {pages} {007}},\ \Eprint {https://arxiv.org/abs/2211.13241} {arXiv:2211.13241 [hep-th]} \BibitemShut {NoStop}%
\bibitem [{\citenamefont {Bossard}\ \emph {et~al.}(2017)\citenamefont {Bossard}, \citenamefont {Cederwall}, \citenamefont {Kleinschmidt}, \citenamefont {Palmkvist},\ and\ \citenamefont {Samtleben}}]{Bossard:2017aae}%
  \BibitemOpen
  \bibfield  {author} {\bibinfo {author} {\bibfnamefont {G.}~\bibnamefont {Bossard}}, \bibinfo {author} {\bibfnamefont {M.}~\bibnamefont {Cederwall}}, \bibinfo {author} {\bibfnamefont {A.}~\bibnamefont {Kleinschmidt}}, \bibinfo {author} {\bibfnamefont {J.}~\bibnamefont {Palmkvist}},\ and\ \bibinfo {author} {\bibfnamefont {H.}~\bibnamefont {Samtleben}},\ }\bibfield  {title} {\bibinfo {title} {{Generalized diffeomorphisms for $E_9$}},\ }\href {https://doi.org/10.1103/PhysRevD.96.106022} {\bibfield  {journal} {\bibinfo  {journal} {Phys. Rev. D}\ }\textbf {\bibinfo {volume} {96}},\ \bibinfo {pages} {106022} (\bibinfo {year} {2017})},\ \Eprint {https://arxiv.org/abs/1708.08936} {arXiv:1708.08936 [hep-th]} \BibitemShut {NoStop}%
\bibitem [{Note1()}]{Note1}%
  \BibitemOpen
  \bibinfo {note} {It is possible to add negative level generators. Due to the Jacobi identity \protect \eqref {eqn:JacobiLieGS}, their structure constants have to be one-cocycles of $G_{\protect \mathrm {S}}$'s Chevalley–Eilenberg complex $\Omega ^\bullet (\protect \mathfrak {g}_{\protect \mathrm {S}},V_i)$ in appropriate representations $V_i$. If they are additionally coboundaries, they can be removed by a constant $G_{\protect \mathrm {D}}$-transformation. Therefore only non-trivial elements of the Lie algebra cohomology $H^1(\protect \mathfrak {g}_{\protect \mathrm {S}},V_i)$ are relevant. For semi-simple $G_{\protect \mathrm {S}}$, there are no such elements according to Whitehead's lemma. Therefore, we set them here to zero.}\BibitemShut {Stop}%
\bibitem [{Note2()}]{Note2}%
  \BibitemOpen
  \bibinfo {note} {In this form, one can easily compute Maurer-Cartan forms and adjoint actions: Assume first $N = e^{\nu }$. Now one can use the identities \begin {equation} \begin {aligned} \protect \mathrm {d}N N^{-1} &= \DOTSB \sum@ \slimits@ _{m=0}^\infty \protect \frac {[ \nu , \protect \mathrm {d}\nu ]_m}{(m+1)!} \hskip 1em\relax \protect \text {and}\hskip 1em\relax \\ N X N^{-1} &= \DOTSB \sum@ \slimits@ _{m=0}^\infty \protect \frac {[ \nu , X ]_m}{m!}\protect \,, \end {aligned} \end {equation} where the bracket $[X, Y]_m$ is defined for $m\ge 0$ as \begin {equation} [X, Y]_m = [ X, [X, Y]_{m-1} ] \hskip 1em\relax \protect \text {with}\hskip 1em\relax [X, Y]_0 = Y \end {equation}}\BibitemShut {NoStop}%
\bibitem [{\citenamefont {Pol\'a\v{c}ek}\ and\ \citenamefont {Siegel}(2014)}]{Polacek:2013nla}%
  \BibitemOpen
  \bibfield  {author} {\bibinfo {author} {\bibfnamefont {M.}~\bibnamefont {Pol\'a\v{c}ek}}\ and\ \bibinfo {author} {\bibfnamefont {W.}~\bibnamefont {Siegel}},\ }\bibfield  {title} {\bibinfo {title} {{Natural curvature for manifest T-duality}},\ }\href {https://doi.org/10.1007/JHEP01(2014)026} {\bibfield  {journal} {\bibinfo  {journal} {JHEP}\ }\textbf {\bibinfo {volume} {01}},\ \bibinfo {pages} {026}},\ \Eprint {https://arxiv.org/abs/1308.6350} {arXiv:1308.6350 [hep-th]} \BibitemShut {NoStop}%
\bibitem [{\citenamefont {Butter}(2022)}]{Butter:2021dtu}%
  \BibitemOpen
  \bibfield  {author} {\bibinfo {author} {\bibfnamefont {D.}~\bibnamefont {Butter}},\ }\bibfield  {title} {\bibinfo {title} {{Exploring the geometry of supersymmetric double field theory}},\ }\href {https://doi.org/10.1007/JHEP01(2022)152} {\bibfield  {journal} {\bibinfo  {journal} {JHEP}\ }\textbf {\bibinfo {volume} {01}},\ \bibinfo {pages} {152}},\ \Eprint {https://arxiv.org/abs/2101.10328} {arXiv:2101.10328 [hep-th]} \BibitemShut {NoStop}%
\bibitem [{\citenamefont {Coimbra}\ \emph {et~al.}(2016)\citenamefont {Coimbra}, \citenamefont {Strickland-Constable},\ and\ \citenamefont {Waldram}}]{Coimbra:2014uxa}%
  \BibitemOpen
  \bibfield  {author} {\bibinfo {author} {\bibfnamefont {A.}~\bibnamefont {Coimbra}}, \bibinfo {author} {\bibfnamefont {C.}~\bibnamefont {Strickland-Constable}},\ and\ \bibinfo {author} {\bibfnamefont {D.}~\bibnamefont {Waldram}},\ }\bibfield  {title} {\bibinfo {title} {{Supersymmetric Backgrounds and Generalised Special Holonomy}},\ }\href {https://doi.org/10.1088/0264-9381/33/12/125026} {\bibfield  {journal} {\bibinfo  {journal} {Class. Quant. Grav.}\ }\textbf {\bibinfo {volume} {33}},\ \bibinfo {pages} {125026} (\bibinfo {year} {2016})},\ \Eprint {https://arxiv.org/abs/1411.5721} {arXiv:1411.5721 [hep-th]} \BibitemShut {NoStop}%
\bibitem [{Note3()}]{Note3}%
  \BibitemOpen
  \bibinfo {note} {This antisymmetrization weights each term with 1/2.}\BibitemShut {Stop}%
\bibitem [{\citenamefont {Grana}\ \emph {et~al.}(2009)\citenamefont {Grana}, \citenamefont {Minasian}, \citenamefont {Petrini},\ and\ \citenamefont {Waldram}}]{Grana:2008yw}%
  \BibitemOpen
  \bibfield  {author} {\bibinfo {author} {\bibfnamefont {M.}~\bibnamefont {Grana}}, \bibinfo {author} {\bibfnamefont {R.}~\bibnamefont {Minasian}}, \bibinfo {author} {\bibfnamefont {M.}~\bibnamefont {Petrini}},\ and\ \bibinfo {author} {\bibfnamefont {D.}~\bibnamefont {Waldram}},\ }\bibfield  {title} {\bibinfo {title} {{T-duality, Generalized Geometry and Non-Geometric Backgrounds}},\ }\href {https://doi.org/10.1088/1126-6708/2009/04/075} {\bibfield  {journal} {\bibinfo  {journal} {JHEP}\ }\textbf {\bibinfo {volume} {04}},\ \bibinfo {pages} {075}},\ \Eprint {https://arxiv.org/abs/0807.4527} {arXiv:0807.4527 [hep-th]} \BibitemShut {NoStop}%
\bibitem [{\citenamefont {Lee}\ \emph {et~al.}(2017)\citenamefont {Lee}, \citenamefont {Strickland-Constable},\ and\ \citenamefont {Waldram}}]{Lee:2014mla}%
  \BibitemOpen
  \bibfield  {author} {\bibinfo {author} {\bibfnamefont {K.}~\bibnamefont {Lee}}, \bibinfo {author} {\bibfnamefont {C.}~\bibnamefont {Strickland-Constable}},\ and\ \bibinfo {author} {\bibfnamefont {D.}~\bibnamefont {Waldram}},\ }\bibfield  {title} {\bibinfo {title} {{Spheres, generalised parallelisability and consistent truncations}},\ }\href {https://doi.org/10.1002/prop.201700048} {\bibfield  {journal} {\bibinfo  {journal} {Fortsch. Phys.}\ }\textbf {\bibinfo {volume} {65}},\ \bibinfo {pages} {1700048} (\bibinfo {year} {2017})},\ \Eprint {https://arxiv.org/abs/1401.3360} {arXiv:1401.3360 [hep-th]} \BibitemShut {NoStop}%
\bibitem [{\citenamefont {Hassler}\ and\ \citenamefont {Sakatani}(2022)}]{Hassler:2022egz}%
  \BibitemOpen
  \bibfield  {author} {\bibinfo {author} {\bibfnamefont {F.}~\bibnamefont {Hassler}}\ and\ \bibinfo {author} {\bibfnamefont {Y.}~\bibnamefont {Sakatani}},\ }\bibfield  {title} {\bibinfo {title} {{All maximal gauged supergravities with uplift}},\ }\bibfield  {journal} {\bibinfo  {journal} {PTEP}\ }\textbf {\bibinfo {volume} {2023}},\ \href {https://doi.org/10.1093/ptep/ptad104} {10.1093/ptep/ptad104} (\bibinfo {year} {2022}),\ \Eprint {https://arxiv.org/abs/2212.14886} {arXiv:2212.14886 [hep-th]} \BibitemShut {NoStop}%
\bibitem [{\citenamefont {Ambrose}\ and\ \citenamefont {Singer}(1958)}]{Ambrose1958}%
  \BibitemOpen
  \bibfield  {author} {\bibinfo {author} {\bibfnamefont {W.}~\bibnamefont {Ambrose}}\ and\ \bibinfo {author} {\bibfnamefont {I.~M.}\ \bibnamefont {Singer}},\ }\bibfield  {title} {\bibinfo {title} {On homogeneous riemannian manifolds},\ }\href@noop {} {\bibfield  {journal} {\bibinfo  {journal} {Duke Math.J.}\ }\textbf {\bibinfo {volume} {25}},\ \bibinfo {pages} {647} (\bibinfo {year} {1958})}\BibitemShut {NoStop}%
\bibitem [{\citenamefont {Klimcik}\ and\ \citenamefont {Severa}(1996)}]{Klimcik:1996np}%
  \BibitemOpen
  \bibfield  {author} {\bibinfo {author} {\bibfnamefont {C.}~\bibnamefont {Klimcik}}\ and\ \bibinfo {author} {\bibfnamefont {P.}~\bibnamefont {Severa}},\ }\bibfield  {title} {\bibinfo {title} {{Dressing cosets}},\ }\href {https://doi.org/10.1016/0370-2693(96)00669-7} {\bibfield  {journal} {\bibinfo  {journal} {Phys. Lett. B}\ }\textbf {\bibinfo {volume} {381}},\ \bibinfo {pages} {56} (\bibinfo {year} {1996})},\ \Eprint {https://arxiv.org/abs/hep-th/9602162} {arXiv:hep-th/9602162} \BibitemShut {NoStop}%
\bibitem [{\citenamefont {Demulder}\ \emph {et~al.}(2020)\citenamefont {Demulder}, \citenamefont {Hassler}, \citenamefont {Piccinini},\ and\ \citenamefont {Thompson}}]{Demulder:2019vvh}%
  \BibitemOpen
  \bibfield  {author} {\bibinfo {author} {\bibfnamefont {S.}~\bibnamefont {Demulder}}, \bibinfo {author} {\bibfnamefont {F.}~\bibnamefont {Hassler}}, \bibinfo {author} {\bibfnamefont {G.}~\bibnamefont {Piccinini}},\ and\ \bibinfo {author} {\bibfnamefont {D.~C.}\ \bibnamefont {Thompson}},\ }\bibfield  {title} {\bibinfo {title} {{Generalised Cosets}},\ }\href {https://doi.org/10.1007/JHEP09(2020)044} {\bibfield  {journal} {\bibinfo  {journal} {JHEP}\ }\textbf {\bibinfo {volume} {09}},\ \bibinfo {pages} {044}},\ \Eprint {https://arxiv.org/abs/1912.11036} {arXiv:1912.11036 [hep-th]} \BibitemShut {NoStop}%
\bibitem [{\citenamefont {Sakatani}(2022)}]{Sakatani:2021skx}%
  \BibitemOpen
  \bibfield  {author} {\bibinfo {author} {\bibfnamefont {Y.}~\bibnamefont {Sakatani}},\ }\bibfield  {title} {\bibinfo {title} {{Poisson\textendash{}Lie T-plurality for dressing cosets}},\ }\href {https://doi.org/10.1093/ptep/ptac079} {\bibfield  {journal} {\bibinfo  {journal} {PTEP}\ }\textbf {\bibinfo {volume} {2022}},\ \bibinfo {pages} {063B01} (\bibinfo {year} {2022})},\ \Eprint {https://arxiv.org/abs/2112.14766} {arXiv:2112.14766 [hep-th]} \BibitemShut {NoStop}%
\bibitem [{\citenamefont {Cassani}\ \emph {et~al.}(2019)\citenamefont {Cassani}, \citenamefont {Josse}, \citenamefont {Petrini},\ and\ \citenamefont {Waldram}}]{Cassani:2019vcl}%
  \BibitemOpen
  \bibfield  {author} {\bibinfo {author} {\bibfnamefont {D.}~\bibnamefont {Cassani}}, \bibinfo {author} {\bibfnamefont {G.}~\bibnamefont {Josse}}, \bibinfo {author} {\bibfnamefont {M.}~\bibnamefont {Petrini}},\ and\ \bibinfo {author} {\bibfnamefont {D.}~\bibnamefont {Waldram}},\ }\bibfield  {title} {\bibinfo {title} {{Systematics of consistent truncations from generalised geometry}},\ }\href {https://doi.org/10.1007/JHEP11(2019)017} {\bibfield  {journal} {\bibinfo  {journal} {JHEP}\ }\textbf {\bibinfo {volume} {11}},\ \bibinfo {pages} {017}},\ \Eprint {https://arxiv.org/abs/1907.06730} {arXiv:1907.06730 [hep-th]} \BibitemShut {NoStop}%
\end{thebibliography}%

\end{document}